\documentclass[aps,prb,twocolumn,superscriptaddress,showpacs]{revtex4-1}
\usepackage{graphicx,bm,mathtools}

\begin{document}

\title{Neutron scattering and muon spin relaxation measurements of the non-centrosymmetric antiferromagnet CeCoGe$_3$ }
\date{\today}

\author{M. Smidman}
\email[]{m.smidman@warwick.ac.uk}
\affiliation{Department of Physics, University of Warwick, Coventry CV4 7AL, United Kingdom}
\author{D. T. Adroja}
\email[]{devashibhai.adroja@stfc.ac.uk}
\affiliation{ISIS Facility, STFC, Rutherford Appleton Laboratory, Chilton, Didcot, Oxfordshire OX11 0QX, United Kingdom}
\author{A. D. Hillier}
\affiliation{ISIS Facility, STFC, Rutherford Appleton Laboratory, Chilton, Didcot, Oxfordshire OX11 0QX, United Kingdom}
\author{L. C. Chapon}
\affiliation{Institut Laue Langevin, BP 156, 38042 Grenoble Cedex 9, France}
\author{J. W. Taylor}
\affiliation{ISIS Facility, STFC, Rutherford Appleton Laboratory, Chilton, Didcot, Oxfordshire OX11 0QX, United Kingdom}
\author{V. K. Anand}
\affiliation{ISIS Facility, STFC, Rutherford Appleton Laboratory, Chilton, Didcot, Oxfordshire OX11 0QX, United Kingdom}
\author{R. P. Singh}
\affiliation{Department of Physics, University of Warwick, Coventry CV4 7AL, United Kingdom}
\author{M. R. Lees}
\affiliation{Department of Physics, University of Warwick, Coventry CV4 7AL, United Kingdom}
\author{E. A. Goremychkin}
\affiliation{ISIS Facility, STFC, Rutherford Appleton Laboratory, Chilton, Didcot, Oxfordshire OX11 0QX, United Kingdom}
\affiliation{School of Physics and Astronomy, University of Southampton, Southampton SO17 1BJ, United Kingdom}
\author{M. M. Koza}
\affiliation{Institut Laue Langevin, BP 156, 38042 Grenoble Cedex 9, France}
\author{V. V. Krishnamurthy}
\affiliation{School of Physics, Astronomy and Computational Sciences, George Mason University, Fairfax, Virginia 22030, USA}
\author{D. M. Paul}
\affiliation{Department of Physics, University of Warwick, Coventry CV4 7AL, United Kingdom}
\author{G. Balakrishnan}
\email[]{g.balakrishnan@warwick.ac.uk}
\affiliation{Department of Physics, University of Warwick, Coventry CV4 7AL, United Kingdom}

\begin{abstract}
The magnetic states of the non-centrosymmetric, pressure induced superconductor CeCoGe$_3$ have been studied with magnetic susceptibility, muon spin relaxation ($\mu$SR), single crystal neutron diffraction and inelastic neutron scattering (INS). CeCoGe$_3$ exhibits three magnetic phase transitions at  $T_{\rm{N1}}$~=~21~K, $T_{\rm{N2}}$~=~12~K and  $T_{\rm{N3}}$~=~8~K. The presence of long range magnetic order below $T_{\rm{N1}}$ is revealed by the observation of oscillations of the asymmetry in the $\mu$SR spectra between 13~K and 20~K and a sharp increase in the muon depolarization rate. Single crystal neutron diffraction measurements reveal magnetic Bragg peaks consistent with propagation vectors of \textbf{k}~=~(0,0,$\frac{2}{3}$) between $T_{\rm{N1}}$ and $T_{\rm{N2}}$, \textbf{k}~=~(0,0,$\frac{5}{8}$) between $T_{\rm{N2}}$ and $T_{\rm{N3}}$ and \textbf{k}~=~(0,0,$\frac{1}{2}$) below $T_{\rm{N3}}$. An increase in intensity of the  (1~1~0) reflection between  $T_{\rm{N1}}$ and $T_{\rm{N3}}$ also indicates a ferromagnetic component in these phases. These measurements are consistent with an equal moment, two-up, two-down magnetic structure below $T_{\rm{N3}}$, with a magnetic moment of 0.405(5)~$\rm{\mu_B}$/Ce. Above $T_{\rm{N2}}$, the results are consistent with an equal moment, two-up, one-down structure with a moment of  0.360(6)~$\rm{\mu_B}$/Ce. INS studies reveal two crystal-field (CEF) excitations at $\sim$~19 and $\sim$~27~meV. From an analysis with a CEF model, the wave-functions of the J~=~$\frac{5}{2}$ multiplet are evaluated along with a prediction for the magnitude and direction of the ground state magnetic moment. Our model correctly predicts that the moments order along the $c$ axis but the observed magnetic moment of 0.405(5)~$\rm{\mu_B}$ is reduced compared to the predicted moment of 1.01~$\rm{\mu_B}$. This is ascribed to hybridization between the localized Ce$^{3+}$ f-electrons and the conduction band. This suggests that CeCoGe$_3$ has a degree of hybridization between that of CeRhGe$_3$ and the non-centrosymmetric superconductor CeRhSi$_3$.  
\end{abstract}


\maketitle

\section{Introduction}

The coexistence of superconductivity (SC) and magnetism in heavy fermion (HF) compounds has attracted considerable research interest recently. In particular, several HF systems appear to exhibit unconventional SC close to a quantum critical point (QCP). On tuning the electronic ground state of these systems by doping, pressure or the application of magnetic fields, the SC appears in regions where the magnetic order is being suppressed. \cite{NatureMagSC,PfleidererRMP} There is great interest therefore in understanding this phenomenon and in particular the role of magnetic fluctuations in potentially mediating the SC of these compounds. Most of the compounds which display HF SC have centrosymmetric crystal structures, in which the Cooper pairs condense in either spin-singlet or spin-triplet states. However, several cerium based compounds with non-centrosymmetric structures have been recently reported to exhibit SC. The first HF NCS reported was CePt$_3$Si, where antiferromagnetic (AFM) order ($T_{\rm{N}}$~=~2.2~K) and SC ($T_{\rm{c}}$ = 0.75 K) coexist at ambient pressure. \cite{CePt3Si2004} In non-centrosymmetric superconductors (NCS), a finite antisymmetric spin-orbit coupling (ASOC) lifts the spin degeneracy of the conduction bands, allowing for the mixture of spin singlet and triplet pairing states. \cite{NCSGorkov} 

We report results of neutron scattering and muon spin relaxation ($\mu$SR) measurements of the NCS CeCoGe$_3$. This is a member of the Ce$TX_3$ (T~=~transition metal, X~=~Si or Ge) series of compounds which crystallize in the non-centrosymmetric, tetragonal BaNiSn$_3$ type structure (space group $I4mm$). In particular, the lack of a mirror plane perpendicular to [0 0 1] leads to a Rashba type ASOC. \cite{BauerNCS} CeCoGe$_3$ orders antiferromagnetically at ambient pressure, with three magnetic phases ($T_{\rm{N1}}$~=~21~K, $T_{\rm{N2}}$~=~12~K, $T_{\rm{N3}}$~=~8~K).\cite{CeCoGe31993,CeCoGe32005} $T_{\rm{N1}}$ decreases as a function of applied pressure and there is an onset of SC for $p~>$~4.3~GPa, with a $T_{\rm{c}}$ of 0.7~K at 5.5~GPa. \cite{CeCoGe3SC} SC is also observed in CeRhSi$_3$ ($p~>$~1.2~GPa) \cite{CeRhSi3SC}, CeIrSi$_3$ ($p~>$~1.8~GPa)  \cite{CeIrSi3SC} and CeIrGe$_3$ ($p~>$~20 GPa). \cite{CeIrGe3SC} The superconducting states of these compounds display highly unconventional properties.  As well as regions of coexistence with AFM order, the upper critical field is highly anisotropic, vastly exceeding the Pauli limiting field along the $c$~axis. \cite{CeCoGe3Hc2} However some members of the Ce$TX_3$ family such as CeCoSi$_3$ and CeRuSi$_3$ do not order magnetically and are intermediate valence compounds. \cite{CeCoGe31998,CeCoSi32007}

The range of observed magnetic properties in the Ce$TX_3$ series has previously been discussed in the context of the Doniach phase diagram \cite{BauerNCS,Doniach1977,HFSC2007,CeTX32008}, with competition between the intersite Ruderman-Kittel-Kasuya-Yosida (RKKY) interaction which favors magnetic ordering and the on-site Kondo effect which leads to a non-magnetic singlet ground state. However, further studies are necessary to characterize the magnetic states of the Ce$TX_3$ series.  Knowledge of the magnetic ground states and crystal electric field (CEF) levels will aid in understanding the relationship between SC and magnetism in the Ce$TX_3$ compounds and allows detailed comparisons between members of the series. In particular, the role of hybridization in determining the phase diagram can be examined. CeCoGe$_3$ can be considered a strongly correlated system with an electronic specific heat coefficient $\gamma~=32~\rm{mJ~mol^{-1}~K^{-2}}$ and an enhanced cyclotron mass of $10 m_{\rm e}$, where $m_{\rm e}$ is the free electron mass. \cite{CeCoGe32005,CeCoGe3FS} The proximity of the compound to quantum criticality has been studied in the  CeCoGe$_{3-x}$Si$_x$ system, where the substitution of Si increases the chemical pressure. Interestingly whilst antiferromagnetism is suppressed for $x~=~1.2$ and a quantum critical region with non-Fermi liquid behaviour is observed for $1~<~x~<~1.5$, no SC was reported down to 0.5~K. \cite{CeCoGe31998,CeCoGe32002} This is in contrast to the superconducting behavior observed for the $x~=~0$ compound with applied hydrostatic pressure. 

As well as being an unconventional superconductor \cite{CeCoGe3Hc2}, CeCoGe$_3$ also has the highest magnetic ordering temperature ($T_{\rm{N1}}$~=~21~K) of any of the Ce$TX_3$ compounds and exhibits a complex temperature-pressure phase diagram. \cite{CeCoGe3HP,CeCoGe3LP} Specific heat measurements of single crystals reveal that under a pressure of $p~=~0.8$~GPa, a fourth transition is observed at 15.3~K in addition to those observed under ambient conditions. \cite{CeCoGe3HP} The temperature of this transition does not shift with pressure whilst $T_{\rm{N1}}$ is suppressed until it meets the pressure induced phase at $p~=~1.5$~GPa. In turn, the transition temperature of this phase is suppressed upon further increasing pressure until it merges with $T_{\rm{N2}}$. The $T~-~P$ phase diagram shows a series of step-like decreases in the magnetic ordering temperature. A total of six phases in the phase diagram were suggested from single crystal measurements up to 7~GPa, whilst eight were observed in polycrystalline samples up to 2~GPa. \cite{CeCoGe3HP} The magnetic order is suppressed at   p~=~5.5~GPa and there is a region of coexistence with SC. The lack of step-like transitions above 3.1~GPa could indicate a change in magnetic structure which may be important for understanding the emergence of SC in the system.

The magnetic structure of CeCoGe$_3$  has previously been studied at ambient pressure using single crystal neutron diffraction in zero field where two propagation vectors were observed at 2.9 K, \textbf{k$_1$} = (0,0,$\frac{1}{2}$) and  \textbf{k$_2$} = (0,0,$\frac{3}{4}$). \cite{CeCoGe3SCND} Powder neutron diffraction measurements also indicate the presence of  \textbf{k$_1$} at 2~K. \cite{CeCoGe3PND} In this study, we have determined the magnetic propagation vector in zero field for each of the three magnetic phases using single crystal neutron diffraction. We are then able to propose magnetic structures for the phases above $T_{\rm{N2}}$ and below $T_{\rm{N3}}$.  We report the temperature dependence of magnetic Bragg reflections from 2~-~35~K. The presence of long range magnetic order is also revealed by $\mu$SR measurements, where oscillations are observed in the spectra below $T_{\rm{N1}}$.  Single crystal susceptibility and magnetization measurements were previously used to suggest a CEF scheme with a ground state doublet consisting of the $ |\pm \frac{1}{2} \rangle$ states. \cite{CeCoGe32005} We use INS to directly measure transitions from the ground state to the excited CEF levels and are able to find an energy level scheme and a set of wave functions compatible with both INS and magnetic susceptibility measurements. We are also able to compare the degree of hybridization in CeCoGe$_3$ with other compounds in the series.

\section{Experimental Details}

Polycrystalline samples of CeCoGe$_3$ and LaCoGe$_3$ were prepared by arc-melting the constituent elements (Ce : 99.99\%, La : 99.99\%, Co : 99.95\%, Ge : 99.999\%) in an argon atmosphere on a water cooled copper hearth.  After being flipped and remelted several times, the boules were wrapped in tantalum foil and annealed at 900~$^{\circ}$C for a week under a dynamic vacuum, better than 10$^{-6}$ Torr. Powder X-ray diffraction measurements were carried out using a Panalytical X-Pert Pro diffractometer. Single crystals were grown by melting polycrystalline material with a bismuth flux following the previously reported technique \cite{CeCoGe32005}.  Plate like single crystals were obtained with faces perpendicular to [0 0 1] and checked using an X-ray Laue imaging system. Excess bismuth was removed by washing the crystals with a solution of 1~:~1  nitric acid.  That the crystals had the correct stoichiometry was confirmed by scanning electron microscopy measurements. Magnetic susceptibility measurements were made using a Quantum Design MPMS SQUID magnetometer. 

Inelastic neutron scattering and $\mu$SR measurements were performed in the ISIS facility at the Rutherford Appleton Laboratory, UK. INS measurements were carried out on the MARI and MERLIN spectrometers. The samples were wrapped in thin Al-foil and mounted inside a thin-walled cylindrical Al-can, which was cooled down to 4.5 K inside a CCR with He-exchange gas around the samples. Incident energies of 10 and 40~meV were used on MARI whilst 15~meV were used on MERLIN, selected via a Fermi chopper. Further low energy INS measurements were carried out on the IN6 spectrometer at the Institut Laue-Langevin, France, with an incident energy of 3.1~meV. $\mu$SR measurements were carried out on the MuSR spectrometer with the detectors in the longitudinal configuration. Spin-polarized muon pulses were implanted into the sample and positrons from the resulting decay were collected in positions either forward or backwards of the initial muon spin direction. The asymmetry is calculated by
\begin{equation}
G_z(t) = \frac{N_F - \alpha N_B}{N_F + \alpha N_B}
\end{equation}
where $N_F$ and $N_B$ are the number of counts at the detectors in the forward and backward positions and $\alpha$ is a constant determined from calibration measurements made in the paramagnetic state with a small applied transverse magnetic field. The maximum asymmetry for an ideal pair of detectors is $\frac{1}{3}$ but this is lower for a real spectrometer. \cite{MuonRef2011} The sample was mounted on a silver plate using GE varnish and cooled in a standard crysotat down to 1.5~K, with He exchange gas around the sample.

Single crystal neutron diffraction measurements were carried out on the D10 instrument at the Institut Laue-Langevin, France. The sample was mounted on an aluminium pin and cooled in a helium-flow cryostat operating down to 2~K. The instrument was operated in the four-circle configuration. An incident wavelength of 2.36~$\rm{\AA}$ was selected using a pyrolytic graphite monochromator. A vertically focused pyrolytic graphite analyzer was used to reduce the background signal. After passing through the analyzer, neutrons were detected using a single $^3$He detector.
\section{Results and discussion}
\begin{figure}[tb]
   \includegraphics[width=0.99\columnwidth]{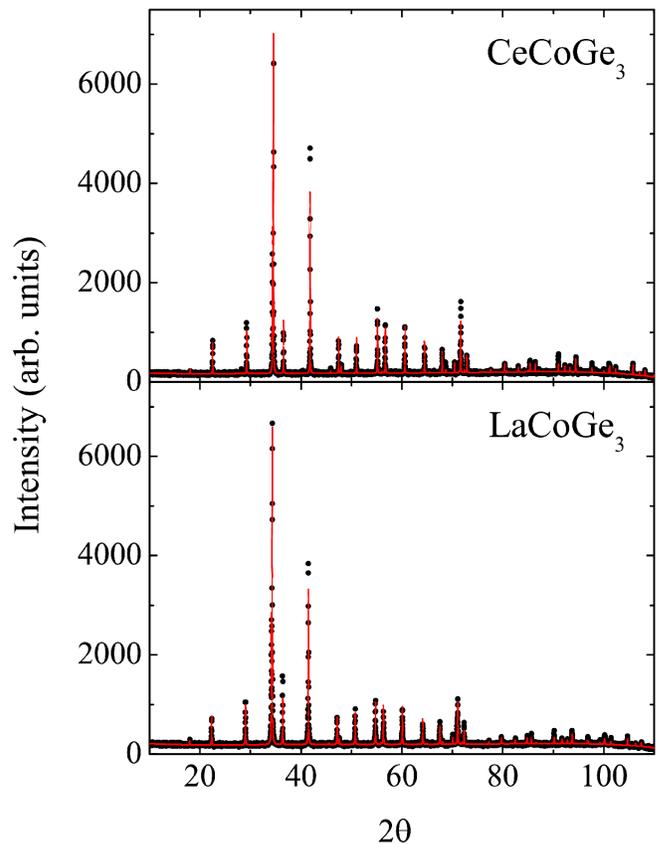}
	\caption{X-ray powder diffraction measurements of polycrystalline CeCoGe$_3$ and LaCoGe$_3$. The solid lines show the Rietveld refinements, the results of which are given in Table \ref{XRDTab}.}
   \label{XRD}
\end{figure}

\begin{table}[ht]
\caption{Results of the refinements of powder x-ray diffraction measurements on CeCoGe$_3$ and LaCoGe$_3$. The lattice parameters, weighted profile factor ($\rm{R_{wp}}$) and the atomic positions are displayed.}
\label{XRDTab} 
\begin{ruledtabular}
 \begin{tabular}{c c c c }

 & CeCoGe$_3$ & LaCoGe$_3$ &  \\
\hline
 $a~(\rm{\AA}$) & 4.32042(4) & 4.35083(7) & \\
 $c~(\rm{\AA}$) &  9.83484(11) & 9.87155(2) & \\
$R_{\rm{wp}}$ & 10.33 & 8.86 & \\
& $x$ & $y$ & $z$ \\
\hline
 Ce &  0 & 0 & 0\\
 Co &  0 & 0 & 0.666(7)\\
 Ge1 &  0 & 0 & 0.4281(6)\\
 Ge2 &  0 & 0.5 & 0.7578(5)\\
 La &  0 & 0 & 0\\
 Co &  0 & 0 & 0.6628(7)\\
 Ge1 &  0 & 0 & 0.4285(6)\\
 Ge2 &  0 & 0.5 & 0.7556(5)\\

\end{tabular}
\end{ruledtabular}
\end{table}

\subsection{Powder X-ray diffraction}
Powder X-ray diffraction measurements were carried out on polycrystalline samples of CeCoGe$_3$ and the isostructural non-magnetic LaCoGe$_3$ at 300~K. A Rietveld refinement was carried out on both samples using the TOPAS software. \cite{TOPAS} The data and refinement are shown in Fig.~\ref{XRD}. One small impurity peak was detectable in CeCoGe$_3$ ($\sim$~1\% of the intensity of the maximum sample peak) whilst none were observed in LaCoGe$_3$, indicating that the samples are very nearly single phase. The site occupancies were all fixed at 100$\%$. The results of the refinements are displayed in Table \ref{XRDTab}. The values of the lattice parameters are in agreement with previously reported values. \cite{CeCoGe31993, CeCoGe3PND}

\subsection{Muon spin relaxation}
\begin{figure}[tb]
   \includegraphics[width=0.8\columnwidth]{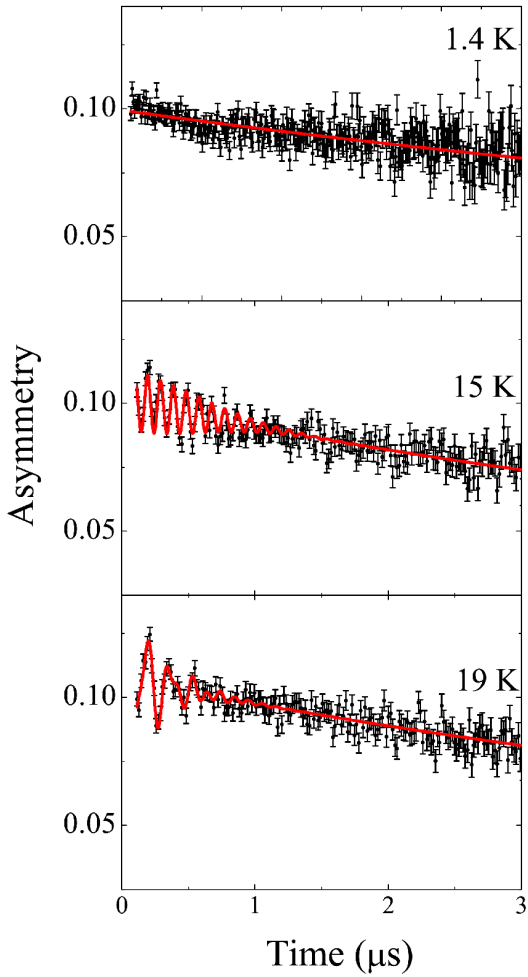}
	\caption{$\mu$SR spectra measured at three temperatures.  At 19~K, two frequencies could be observed whilst at 15~K only one frequency was observed.  At 1.4~K no oscillations in the spectra were observed.  The solid lines show the fits as described in the text.}
   \label{specT}
\end{figure}
\begin{figure}[tb]
   \includegraphics[width=0.8\columnwidth]{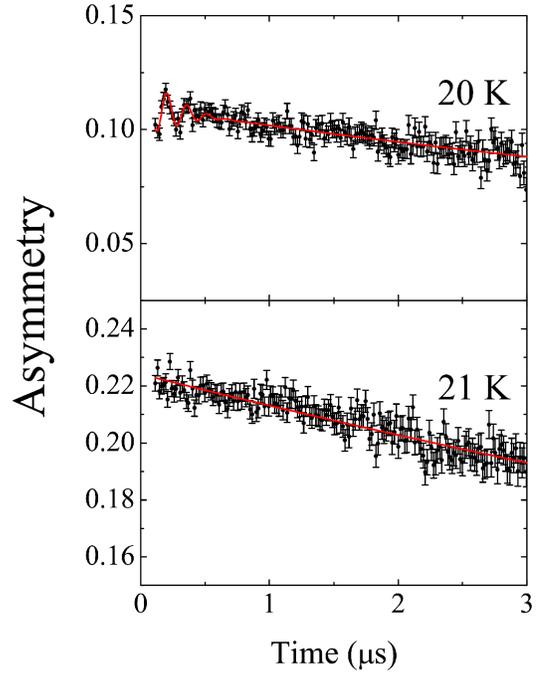}
	\caption{$\mu$SR spectra measured at 20 and 21~K. At 20~K one frequency is observed in the spectrum and the initial asymmetry is reduced whilst at 21~K no oscillations are observed and the initial asymmetry reaches the full value for the instrument. The solid lines show the fits as described in the text.}
   \label{specTn}
\end{figure}
\begin{figure}[tb]
   \includegraphics[width=0.99\columnwidth]{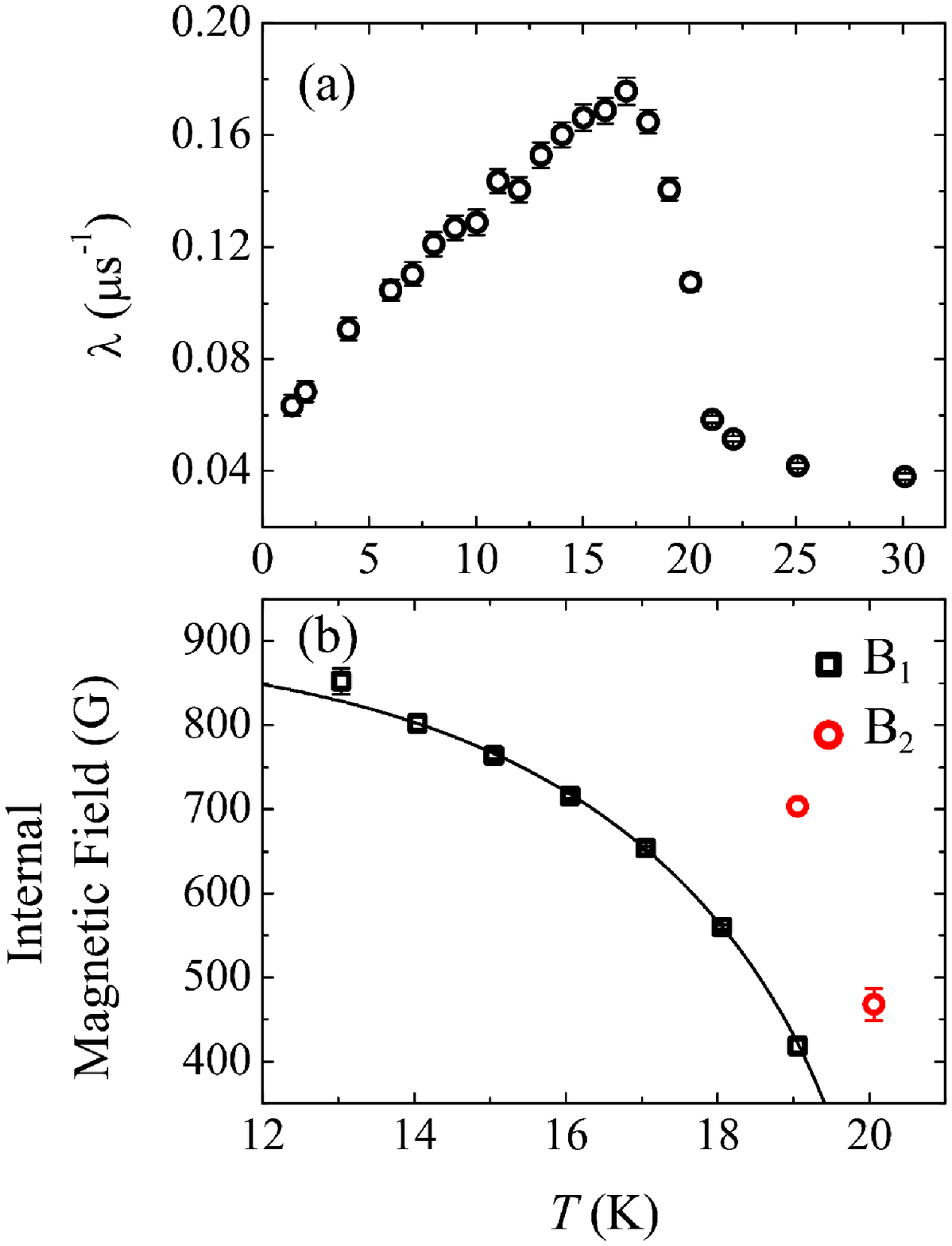}
	\caption{(a) The muon depolarization rate as a function of temperature. (b) The internal fields deduced from the frequencies of the oscillations observed in zero-field $\mu$SR spectra. The solid curve is a fit of $B_1$ to a mean field model described in the text.}
   \label{BFields}
\end{figure}
To investigate the nature of magnetic ordering in CeCoGe$_3$, we measured the zero-field muon spin relaxation of a polycrystalline sample. In the range 13~K~$<$~$T$~$<$~20~K, oscillations of the asymmetry are observed in the $\mu$SR spectra, indicating the presence of long-range magnetic order (Fig.~\ref{specT} and \ref{specTn}). The presence of an oscillation at 20~K (Fig.~\ref{specTn}) as well as a reduced initial asymmetry indicates that the system is ordered at 20~K. However at 21~K, no oscillations are observed and the initial asymmetry reaches the full value for the instrument indicating that $T_{\rm{N1}}$ lies between 20 and 21~K. The spectra were fitted with 
\begin{equation}
G_z(t) = \sum_{i=1}^{n} A_i {\rm{cos(\gamma_\mu}} B_i t+\phi)e^{- \frac{(\sigma_i t)^2}{2}} + A_0e^{-\lambda t} + A_{\rm{bg}}
\end{equation}
where $A_{\rm{i}}$ are the amplitudes of the oscillatory component, $A_0$ is the initial amplitude of the exponential decay,  $B_{\rm{i}}$ are the magnetic fields at the muon site $\rm{i}$, $\sigma_{\rm{i}}$ is the Gaussian decay rate, $\lambda_{\rm{i}}$ is the muon depolarization rate, $\phi$ is the common phase, $\rm{{\gamma_\mu}}/{2\pi}$~=~135.53 MHz~T$^{-1}$ and $A_{\rm{bg}}$ is the background. All the oscillatory spectra could be fitted with one internal magnetic field ($n~=~1$) apart from at 19~K when it was fitted with two internal magnetic fields ($n~=~2$). This implies that there are at least two muon sites but below 19~K it is likely that B$_2$ exceeds the maximum internal field detectable on the MuSR spectrometer due to the pulse width of the ISIS muon beam. Below 13~K the spectra were fitted with just an exponential decay term. The temperature dependence of one of the internal fields was fitted with 
\begin{equation}
B(T) = B(0)\left(1-\left(\frac{T}{T_N}\right)^\alpha \right)^\beta
\end{equation}
With  $\beta$ fixed at 0.5 for a mean field magnet, values of $B(0)$~=~889(16)~G, $\alpha$~=~4.7(4) and $T_{\rm{N}}$~=~20.12(8)~K were obtained (Fig.~\ref{BFields}). A good fit with $\beta$~=~0.5 means the observations are consistent with that of a mean field magnet. The large value of $\alpha$ indicates complex interactions between the magnetic moments. It was also possible to fit the data with $\beta$~=~0.367 and 0.326 for a 3D Heisenberg and Ising model respectively. \cite{Blundell} However, fits with both these values of $\beta$ gave values of $T_{\rm{N}}~<~20$~K and poor fits were obtained for $T_{\rm{N}}~>~20$~K. Since the presence of long-range magnetic order has been observed at 20~K (Fig.~\ref{specTn}), the data are incompatible with these models. The muon depolarization rate ($\lambda$) was found to suddenly increase at $T_{\rm{N1}}$, indicating a transition between the paramagnetic and ordered states. However $\lambda$ does not show a significant anomaly at either $T_{\rm{N2}}$ or $T_{\rm{N3}}$ where there is a rearrangment of the spins and a change in the magnetic structure.
\begin{figure}[tb]
   \includegraphics[width=0.99\columnwidth]{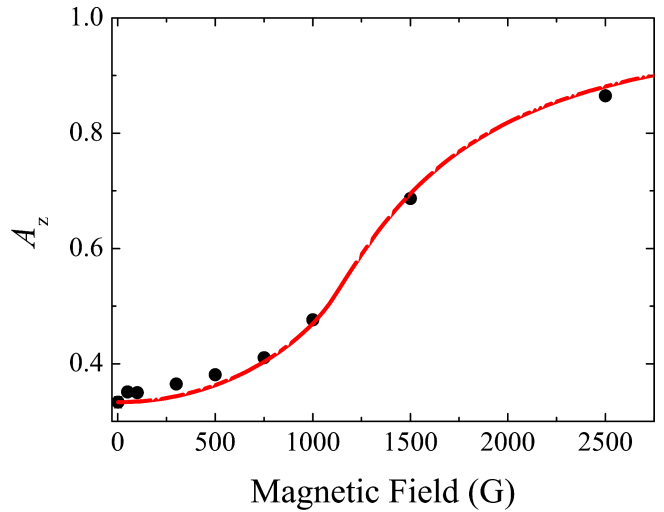}
	\caption{The normalized longitudinal component of the initial asymmetry ($A_{\rm{z}}$) as a function of an applied magnetic field at 1.4~K. The solid line shows a fit described in the text.}
   \label{LFmuSR}
\end{figure}
The initial value of the asymmetry ($A_{\rm{z}}$) as a function of applied longitudinal field at 1.4~K is shown in Fig.~\ref{LFmuSR}. This is the longitudinal component and has been normalized such that $A_{\rm{z}}~=~1$ corresponds to the muon being fully decoupled from its local environment. A fit has been made using the expression described in Ref. \onlinecite{PrattMuSR}. An internal field of 1080(40)~G was obtained which is in approximate agreement with that deduced from the zero field data, despite a change in magnetic structure between 13~K and 1.4~K.

\subsection{Single crystal neutron diffraction}
\begin{figure}[tb]
   \includegraphics[width=0.99\columnwidth]{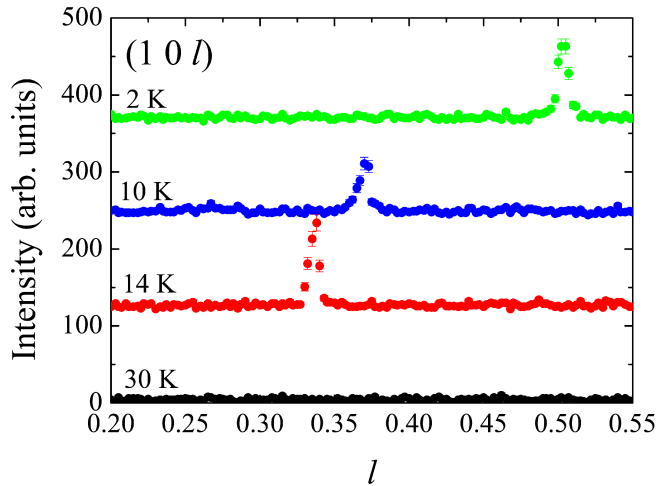}
	\caption{Elastic scans made across (1~0~$l$) at four temperatures. No peak is observed above $T_{\rm{N1}}$. Below 2~K  a peak is observed at $l~=~\frac{1}{2}$, which shifts to $l~=~\frac{3}{8}$ at 10~K and $l~=~\frac{1}{3}$ at 14~K. }
   \label{hklscan}
\end{figure}
\begin{figure}[tb]
   \includegraphics[width=0.99\columnwidth]{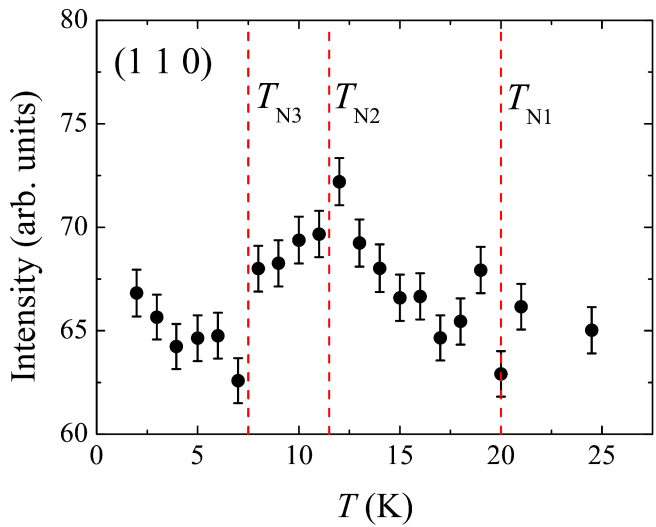}
	\caption{The temperature dependence integrated intensity of the (1~1~0) reflection. An increase in the intensity between $T_{\rm{N3}}$ and $T_{\rm{N1}}$ indicates there is a ferromagnetic contribution in these phases.}
   \label{110peak}
\end{figure}
\begin{figure}[tb]
   \includegraphics[width=0.99\columnwidth]{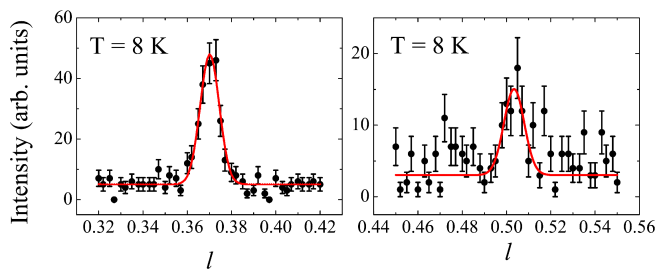}
	\caption{Elastic scans made across made across (1~0~$l$) at 8~K. At this temperature there is a coexistence between the peaks at $l~=~\frac{3}{8}$ and  $l~=~\frac{1}{2}$.}
   \label{8Kpeaks}
\end{figure}

Single crystal neutron diffraction measurements were carried out in each of the three magnetically ordered phases, on the D10 diffractometer . Fig.~\ref{hklscan} shows elastic scans made across (1~0~$l$) at different temperatures.  This reveals that below 20~K, additional peaks for non-integer $l$ are observed, indicating the onset of antiferromagnetic ordering. At 2~K the additional peak is at $l~=~\frac{1}{2}$, at 10~K it is at $l~=~\frac{3}{8}$ and at 14~K it is at $l~=~\frac{1}{3}$. Since the (1~0~0) peak is forbidden for a body-centred structure, this indicates a propagation vector of  \textbf{k}~=~(0,0,$\frac{1}{2}$) below $T_{\rm{N3}}$, \textbf{k}~=~(0,0,$\frac{5}{8}$) for $T_{\rm{N3}}~<~T~<~T_{\rm{N2}}$, and \textbf{k}~=~(0,0,$\frac{2}{3}$) for $T_{\rm{N2}}~<~T<~T_{\rm{N1}}$. Fig. \ref{110peak} shows the intensity of the (1~1~0) reflection between 2 and 25~K.  The increase in integrated intensity of this nuclear peak for $T_{\rm{N3}}~<~T~<~T_{\rm{N1}}$ indicates the presence of an additional ferromagnetic (FM) component for these two magnetic phases. The propagation vector of \textbf{k}~=~(0,0,$\frac{1}{2}$) agrees with the previous single crystal  neutron diffraction measurements. \cite{CeCoGe3SCND} However as shown in Fig.~\ref{hklscan} we do not see a peak at (1~0~$\frac{1}{4}$) as previously observed nor do we observe any evidence for a two component magnetic structure. However at 8~K, just above  $T_{\rm{N3}}$, coexistence of the (1~0~$\frac{1}{2}$) and (1~0~$\frac{3}{8}$) reflections are observed (Fig.~\ref{8Kpeaks}), indicating a first-order transition between the phases. This is also supported by the observation of hysteresis in magnetic isotherms at 3~K.\cite{CeCoGe31993}

\begin{figure}[tb]
   \includegraphics[width=0.99\columnwidth]{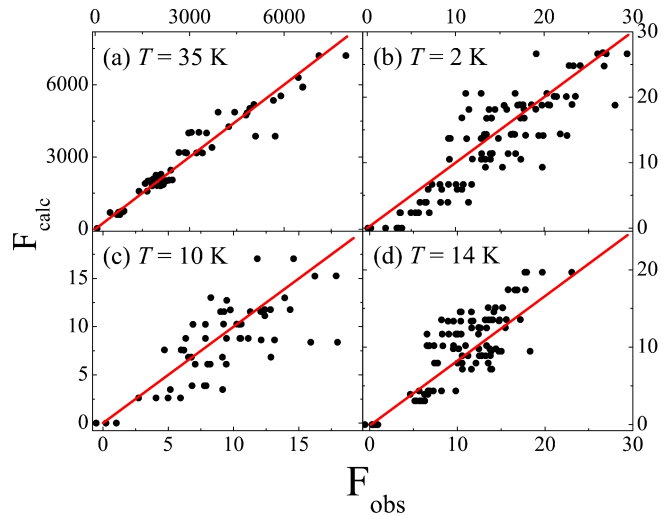}
	\caption{Plots of the calculated vs observed values of F$_{\rm{hkl}}$ for the refinement of  (a) the crystal structure at 35~K and  (b)~-~(d) the magnetic structure in the three magnetic phases. The solid lines indicate where F$_{\rm{calc}}$~=~F$_{\rm{obs}}$.}
   \label{Fcalc}
\end{figure}

\begin{figure}[tb]
   \includegraphics[width=0.99\columnwidth]{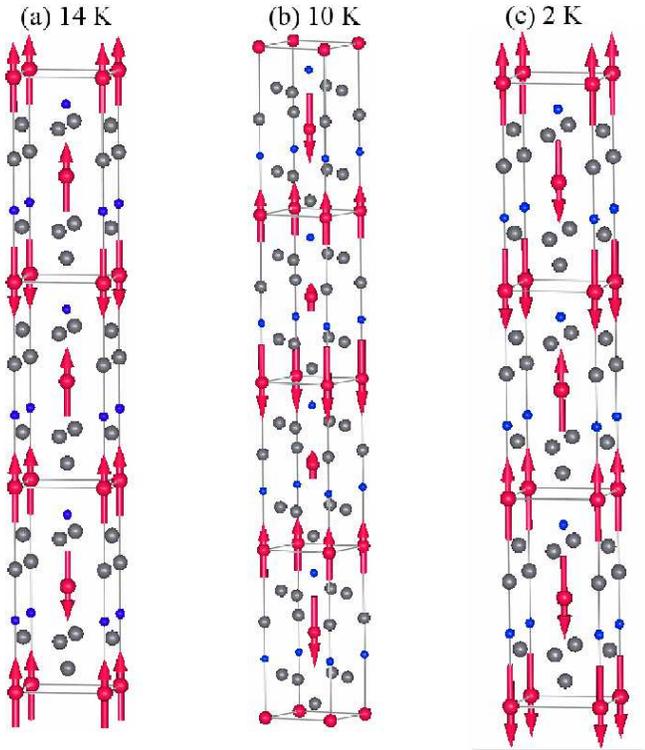}
	\caption{(Color online) The crystal structure of CeCoGe$_3$ where the Ce atoms are in red, the Co in blue and the Ge in grey.  The arrows depict the magnetic moments on the Ce atoms. (a) The proposed magnetic structure at 14~K consisting of the antiferromagnetic component with a global phase $\phi$~=~0 and a ferromagnetic component to give an equal moment, two-up, one-down structure. (b) The antiferromagnet component ($\phi$~=~0) at 10~K for one half of the magnetic unit cell. (c) The magnetic structure at 2~K, with $\phi$~=~$\pi$/4 to give an equal moment, two-up, two down structure.}
   \label{CrysStruc}
\end{figure}

At 35~K, in the paramagnetic state, the intensities were collected for all the allowed, experimentally accessible reflections ($h~k~l$). In each magnetic phase, intensities were collected for the reflections ($h~k~l$)~$\pm$ \textbf{k}. The intensities of 104 magnetic reflections were collected at 2 and 14~K whilst 57 were collected at 10~K. No magnetic peaks were observed corresponding to (0~0~$l$), indicating that in all three phases the magnetic moments point along the $c$ axis. A symmetry analysis of each phase using SARA$h$ \cite{Sarah} shows that $\rm{\Gamma_2}$ is the only irreducible representation of the little group (G$_k$) with the moments along the $c$~axis. Both the crystal and magnetic structures of each phase were fitted using FullProf. \cite{Fullprof1} With the scale factor and extinction parameters fixed from the results of the crystal structure refinement, the only free parameter in the refinements of the magnetic phases was the magnetic moment on the Ce atoms. An R factor of 10.9 was obtained for the refinement of the crystal structure, 21.5 for the magnetic phase at 2~K, 24.3 at 10~K and 22 at 14~K. Plots of  F$_{\rm{calc}}$~vs~F$_{\rm{obs}}$ for all the refinements are shown in Fig.~\ref{Fcalc}.
The introduction of a global phase $\phi$ to a magnetic structure leaves the neutron diffraction pattern unchanged. However for the phase at 2~K with \textbf{k}~=~(0,0,$\frac{1}{2}$), selecting $\phi$~=~$\pi$/4 gives an equal moment on each Ce site of 0.405(5)~$\rm{\mu_B}$. This structure has a two-up, two-down configuration along the $c$~axis (Fig.~\ref{CrysStruc}(c)). Similarly for the phase at 14~K with \textbf{k}~=~(0,0,$\frac{2}{3}$), selecting $\phi$~=~0 gives a modulated structure along the $c$~axis with an up moment of 0.485(6)~$\rm{\mu_B}$ followed by two down moments of 0.243(3)~$\rm{\mu_B}$. The addition of a FM component of  $-$0.125~$\rm{\mu_B}$/Ce gives a constant moment, two-up, one-down configuration as shown in Fig.~\ref{CrysStruc}(a). A FM component is observed in this phase (Fig.~\ref{110peak}) and this equal moment solution is compatible with magnetization results. \cite{CeCoGe32005} For the phase at 10~K with  \textbf{k}~=~(0,0,$\frac{5}{8}$), we were unable to deduce a global phase $\phi$ to which a FM component could be added to give an equal moment solution. A simple three-up, one down structure as was previously suggested for this phase from magnetization measurements  \cite{CeCoGe32005} is not compatible with this propagation vector. The antiferromagnetic component with $\phi$~=~0 is shown in Fig.~\ref{CrysStruc}(b) for half of the magnetic unit cell. However as shown in Fig.~\ref{110peak}, there is also a ferromagnetic component in this phase and further measurements of the nuclear reflections at 10~K would be required to determine the size of this contribution. 

\subsection{Inelastic neutron scattering}

\begin{figure}[tb]
   \includegraphics[width=0.99\columnwidth]{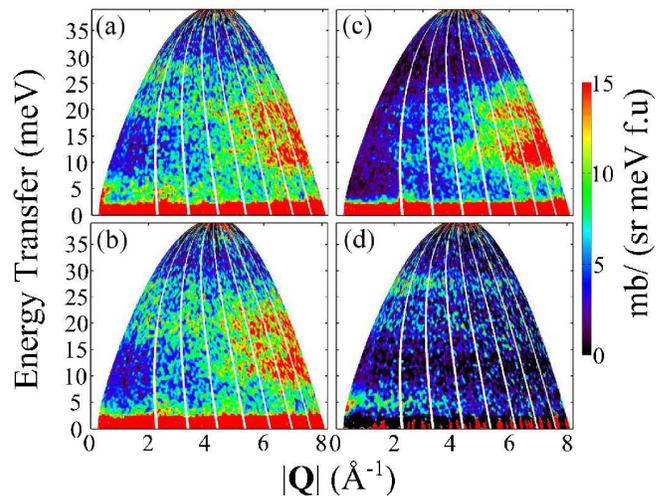}
	\caption{(Color online) Color coded plots of the inelastic neutron scattering intensity with an incident energy of 40 meV (in units of mb~sr$^{-1}$~meV$^{-1}$~f.u$^{-1}$) for (a) CeCoGe$_3$ at 4~K, (b) CeCoGe$_3$ at 25~K and (c) LaCoGe$_3$ at 5~K. The magnetic scattering of CeCoGe$_3$ at 4~K obtained by subtracting the phonon contribution of CeCoGe$_3$ (see text) is shown in (d).}
   \label{colourplots}
\end{figure}
\begin{figure}[tb]
   \includegraphics[width=0.99\columnwidth]{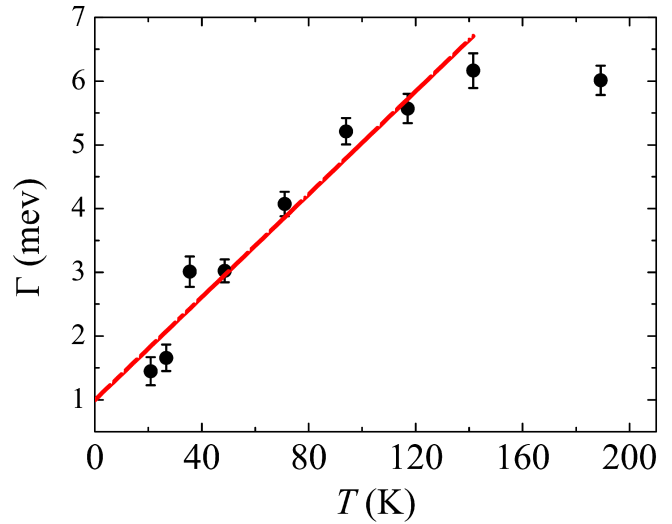}
	\caption{The temperature dependence of the quasielastic linewidth (HWHM) obtained from fitting data measured with an incident energy of 15~meV (see text). A linear fit of the data between 20~K and 150~K is displayed.}
   \label{GammaT}
\end{figure}
\begin{figure}[tb]
   \includegraphics[width=0.99\columnwidth]{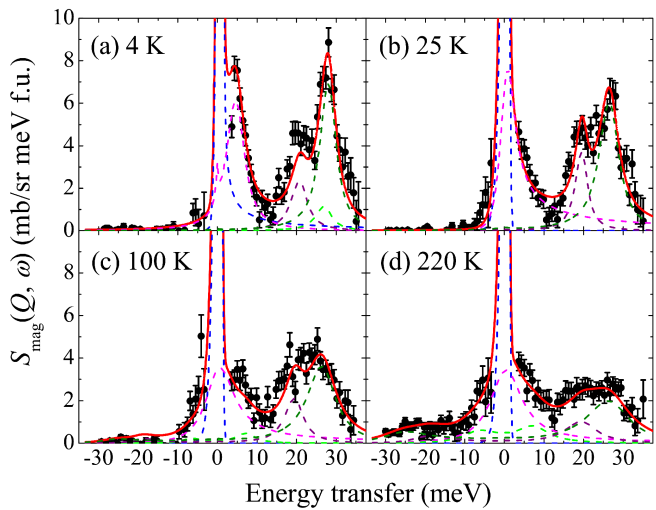}
	\caption{(Color online) Cuts of $S_{\rm{mag}} (Q, \omega)$ with an incident energy of 40~meV integrated over \textbf{$|$Q$|$} from 0 to 3~$\rm{\AA^{-1}}$. The solid lines show fits made to a CEF model described in the text. The components of the fits are shown with dashed lines.}
   \label{40meVFits}
\end{figure}

\begin{figure}[tb]
   \includegraphics[width=0.65\columnwidth]{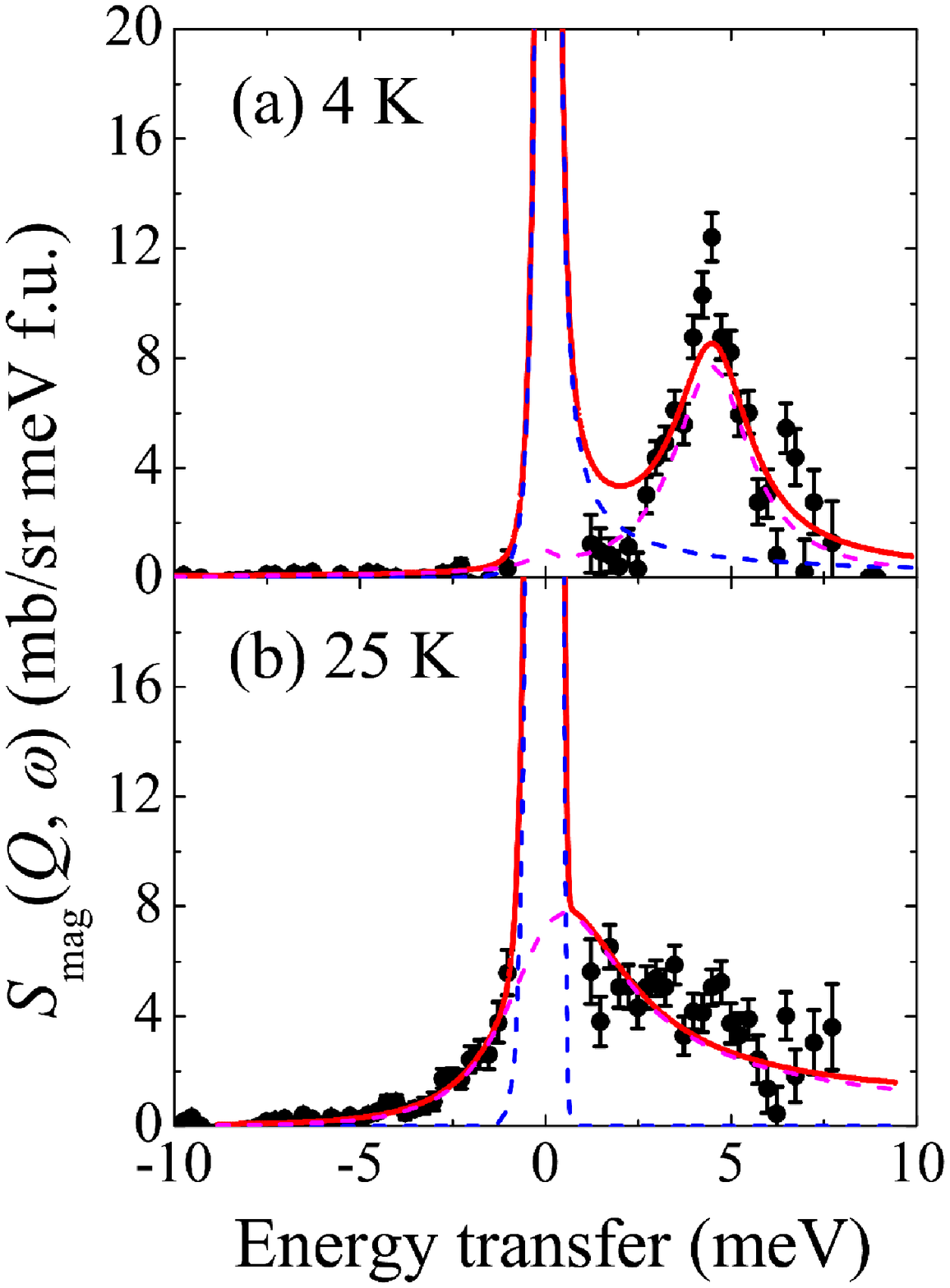}
	\caption{(Color online) Cuts of $S_{\rm{mag}} (Q, \omega)$ with an incident energy of 10~meV integrated over \textbf{$|$Q$|$} from 0 to 2~$\rm{\AA^{-1}}$. Fits are made to a CEF model (see text). The components of the fits are shown with dashed lines. }
   \label{10meVFits}
\end{figure}

\begin{figure}[tb]
   \includegraphics[width=0.99\columnwidth]{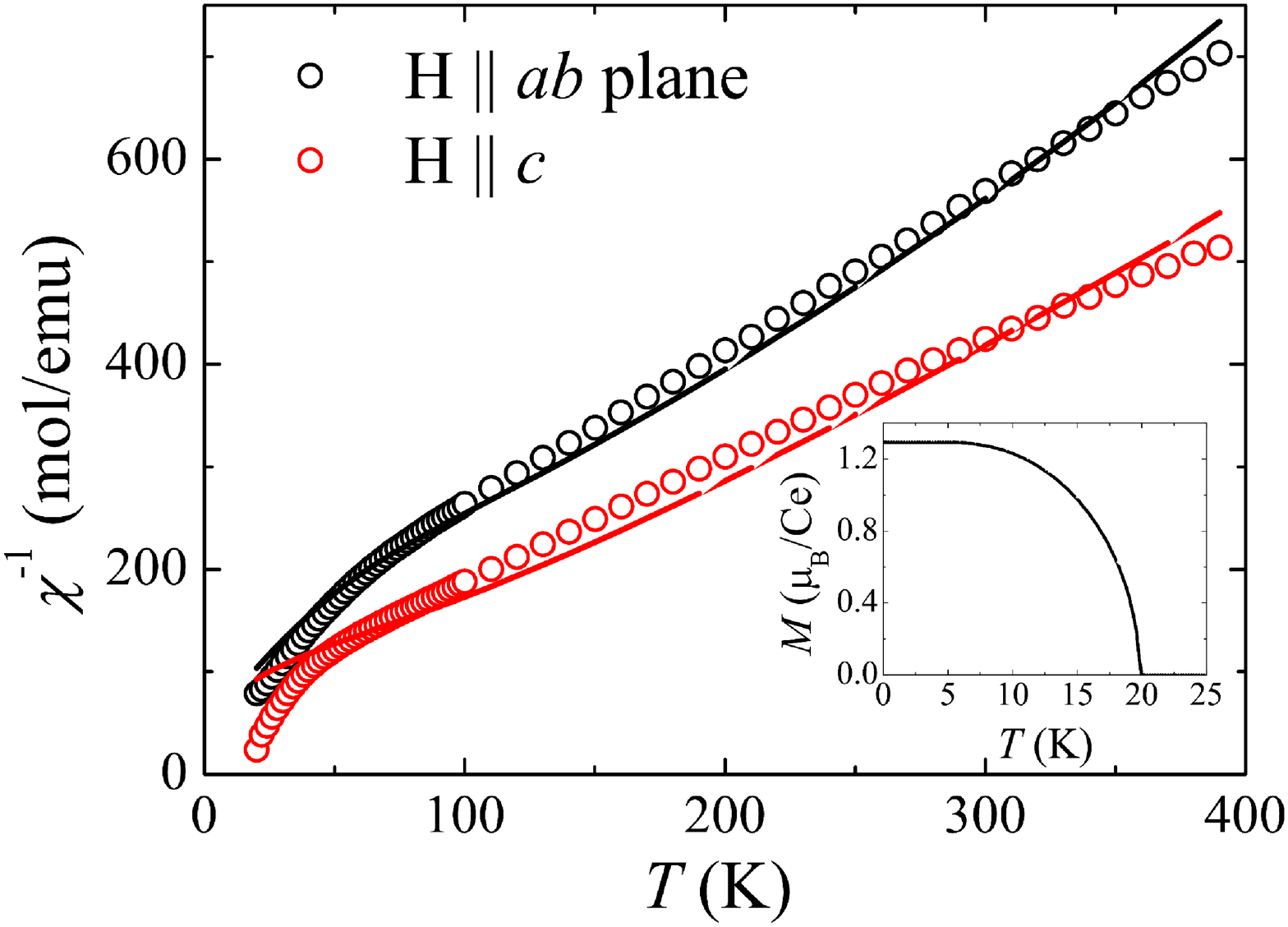}
	\caption{(Color online) The single crystal susceptibility between 20 and 390~K with an applied field of 1000~G. The solid lines show fits to a CEF model (see text). The CEF parameters were fixed from the INS data but anisotropic molecular fields ($\lambda_{ab}$ and $\lambda_{c}$) and temperature independent susceptibilities were fitted. The inset shows a self-consistent mean field calculation of the magnetization per cerium atom using the fitted CEF parameters and a molecular field parameter of 38.9~mol/emu.}
   \label{ChiT}
\end{figure}

To obtain information about the CEF scheme and the magnetic excitations of the ordered state, INS measurements were carried out on polycrystalline samples of CeCoGe$_3$ and LaCoGe$_3$ using the MARI spectrometer with incident neutron energies (E$_i$) of 10 and 40~meV. LaCoGe$_3$ is non-magnetic and isostructural to CeCoGe$_3$ and the measurements were used to estimate the phonon contribution to the scattering. Color coded plots of the INS intensity of CeCoGe$_3$ are shown in Fig.~\ref{colourplots}(a) and \ref{colourplots}(b) at 4 and 25~K respectively, whilst the scattering of LaCoGe$_3$ is shown Fig.~\ref{colourplots}(c).  In both the magnetically ordered and paramagnetic states, two inelastic excitations are observed with a significant intensity at low scattering vectors ($\textbf{Q}$). These are absent in the scattering of non-magnetic LaCoGe$_3$, indicating they are magnetic in origin. The excitations have a maximum intensity at approximately 19 and 27~meV. These can be seen in Fig.~\ref{colourplots}(d) which shows the magnetic scattering ($S_{\rm{mag}} (Q, \omega)$)  obtained from $S_{\rm{Ce}}(Q, \omega)$~-~$\rm{\alpha}$~$S_{\rm{La}}(Q, \omega)$, where $\alpha$~=~0.9, the ratio of the scattering cross sections of CeCoGe$_3$ and LaCoGe$_3$. The scattering intensity decreases with $|\textbf{Q}|$, as expected for CEF excitations. The presence of two CEF excitations is expected for a Ce$\rm^{3+}$ ion in a tetragonal CEF, since according to Kramers theorem, provided time reversal symmetry is preserved, the energy levels of a system with an odd number of electrons, must remain doubly degenerate. Therefore the 6-fold $J~=~\frac{5}{2}$ ground state can be split into a maximum of three doublets in the paramagnetic state.

Also revealed in the 4~K data is an additional excitation with a maximum at around 4.5~meV. This excitation is not present at 25~K (Fig.~\ref{colourplots}(b)), where instead the elastic line is broader.  This indicates the presence of spin waves in the ordered state at 4~K with an energy scale of approximately 4.5~meV for the zone boundary magnons. Interestingly the spin wave peak in CeRhGe$_3$ is observed at around 3~meV and the compound orders at $T_{\rm{N1}}$~=~14.5~K.  \cite{CeRhGe32012} Therefore the spin wave energy appears to similarly scale with $T_{\rm{N1}}$ in both CeRhGe$_3$ and CeCoGe$_3$. Additional low energy measurements on IN6 with an incident energy of 3.1~meV display a lack of magnetic scattering below 2~meV at 4~K, indicating a spin gap in the magnon spectrum. In the paramagnetic state, the spectral weight is shifted towards the elastic line and quasielastic scattering (QES) is observed. This is additional magnetic scattering, centred on the elastic line but with a linewidth broader than the instrument resolution. Further measurements were made in the paramagnetic state between 20 and 200~K on the MERLIN spectrometer with an incident energy of 15~meV. The temperature dependence of the half width at half maximum ($\Gamma$) is shown in Fig.~\ref{GammaT}. The data were fitted with an elastic line resolution function and an additional Lorentzian function to model the quasielastic component. The widths of the elastic component were fixed from measurements of vanadium with the same incident energy and frequency of the Fermi chopper. An estimate of the Kondo temperature ($T_{\rm{K}}$) can be obtained from the value of $\Gamma$ at 0~K. From a linear fit to the data we estimate $T_{\rm{K}}$~=~11(3)~K. This is of the same order as the ordering temperature $T_{\rm{N1}}$~=~21~K. A linear dependence of the QES linewidth with temperature is expected until the thermal energy approaches the splitting of the first excited CEF level. \cite{Linewidth} The first CEF excitation is at  19~meV~($\sim$220~K), which may explain the deviation from linear behaviour observed at 190~K.  It was also possible to fit the data to a $T^{\frac{1}{2}}$ dependence. This behaviour has been observed in the linewidth of the QES scattering in other HF systems. \cite{LinewidthHalf}  However this fit yields a negative value of $\Gamma(0)$ for which we have no physical interpretation and therefore has not been displayed.

Cuts of $S_{\rm{mag}} (Q, \omega)$ were made by integrating across low values of \textbf{$|$Q$|$} (0 to 3 $\rm{\AA^{-1}}$). These are shown for $E_{\rm{i}}$~=~40~meV in Fig.~\ref{40meVFits} and for $E_{\rm{i}}$~=~10~meV in Fig.~\ref{10meVFits}. The data were analyzed with the following Hamiltonian for a Ce$\rm^{3+}$ ion at a site with tetragonal point symmetry:
\begin{equation}
\mathcal{H}_{\rm{CF}} = B_2^0{\rm{O_2^0}} + B_4^0{\rm{O_4^0}} + B_4^4{\rm{O_4^4}}
\end{equation}
where $B_n^m$ are CEF parameters and $\rm{O_n^m}$ are the Stevens operator equivalents. Using the fact that Stevens operator equivalents can be expressed in terms of angular momentum operators, the CEF wavefunctions and energies may be determined from diagonalizing $\mathcal{H}_{\rm{CF}}$. \cite{StevensCEF,HutchingsCEF} We sought to find a CEF scheme compatible with both INS and magnetic susceptibility data. $B_2^0$ can be estimated for isotropic exchange interactions, from the high temperature magnetic susceptibility \cite{Jensenrare} using the relation:
\begin{equation}
B_2^0 = \frac{10k_B(\theta_{ab} - \theta_c)}{3(2J -1)(2J +3)}
\end{equation}
where $\theta_{ab}$ and $\theta_{c}$ are the Curie-Weiss temperatures for fields applied in the $ab$~plane and along the $c$~axis respectively. Using the previously obtained values \cite{CeCoGe32005}, $B_2^0$ is calculated to be $-$0.376~meV.  In particular, since $\theta_{ab}$~$<$~$\theta_{c}$, a negative $B_2^0$ is anticipated. We then fitted the INS data in the paramagnetic state with $E_i$~=~10 and 40 meV to obtain values of $B_n^m$. Initially we fixed  $B_2^0$~=~-0.376~meV and varied $B_4^0$ and $B_4^4$. In the final fit, all three CEF parameters were varied. The fits are shown in Figs.~\ref{40meVFits}(b)-(d) and \ref{10meVFits}(b) and it can be seen that there is a good fit to the INS data. Using these values of $B_n^m$, a fit was made to the single crystal susceptibility data, which shows reasonably good agreement (Fig.~\ref{ChiT}). Simultaneously fitting the magnetic susceptibility and the INS data at 25~K, led to similar values of $B_n^m$. At 4~K, in the ordered state, an additional peak is observed in $S_{\rm{mag}} (Q, \omega)$ at around 4.5~meV. Although the full treatment of this data would require a calculation of the spin-wave excitations, we sought to determine if the addition of an internal magnetic field could satisfactorily account for this peak in the ordered state. Since the magnetic moments lie along the $c$~axis below $T_{\rm{N1}}$, we fitted $S_{\rm{mag}}(Q, \omega)$ with a finite internal field $B_{\rm{z}}$, allowing $B_4^0$ and $B_4^4$ to vary. A small change in the CEF parameters was allowed below $T_{\rm{N}}$. This is expected due to small changes in the lattice parameters upon magnetic ordering. As shown in Fig~\ref{40meVFits}.(a) and Fig~\ref{10meVFits}.(a), a $B_{\rm{z}}$ of 1.69(9)~meV gives a good fit to the data. The resulting CEF parameters are shown in Table \ref{CEFTab}. The wavefunctions calculated for the paramagnetic state are 
\begin{equation}
\begin{multlined}
|\psi_1^{\pm} \rangle = 0.8185\left|\pm \frac{5}{2} \right \rangle - 0.5745\left|\mp \frac{3}{2} \right \rangle  \\ \\      
|\psi_2^{\pm} \rangle = \left|\pm \frac{1}{2} \right \rangle \\ \\
|\psi_3^{\pm} \rangle = 0.8185\left|\pm \frac{3}{2} \right \rangle + 0.5745\left|\mp \frac{5}{2}  \right \rangle
\end{multlined}
\end{equation}
$\psi_1 (\Gamma_6(1))$ is predicted to be the ground state (GS) wavefunction whilst $\psi_2 (\Gamma_7)$ is 19.3 meV and $\psi_3 (\Gamma_6(2))$ is 26.4 meV above the GS. The GS magnetic moments of the cerium atoms in the ab-plane ($\langle \mu_x \rangle$) and along the $c$ axis ($\langle \mu_z \rangle$)  can be calculated from
\begin{equation}
\begin{multlined}
\langle \mu_z \rangle =  \langle \psi_1^{\pm} | g_JJ_z  | \psi_1^{\pm}  \rangle \\ \\
\langle \mu_x \rangle = \langle \psi_1^{\mp}| \frac{g_J}{2}(J^+ + J^-) \left | \psi_1^{\pm} \right \rangle 
\end{multlined}
\end{equation}

The magnitude of $\langle \mu_z \rangle$ is calculated to be 1.01~$\rm{\mu_B}$ whilst the magnitude of $\langle \mu_x \rangle$ is calculated to be 0.9~$\rm{\mu_B}$. A self-consistent mean field calculation of the magnetization shown in the inset of Fig. \ref{ChiT}, gives a ground state magnetic moment of  1.3~$\rm{\mu_B}$. A molecular field parameter of $\lambda~=~38.9~$mol/emu was chosen to correctly reproduce the observed value of $T_{\rm{N1}}$ and this is in good agreement with the values shown in Table~\ref{CEFTab}. However the refinement of the single crystal neutron diffraction data at 2~K predicts a moment along the $c$ axis of 0.405(5)~$\rm{\mu_B}$. This implies there is a reduction in the cerium moment due to hybridization between the GS and the conduction electrons. By considering the magnetocrystalline anisotropy energy ($E_{\rm{a}}$), the moment is predicted to lie along the $c$~axis for a negative $B_2^0$ and the $\psi_1$ GS. \cite{MagAniso1990} Therefore our CEF model correctly predicts the direction of the observed magnetic moment. From previous studies of the magnetic susceptibility, a CEF scheme with a GS of $|\pm \frac{1}{2} \rangle$ was suggested. \cite{CeCoGe32005} These CEF parameters, give rise to energy level splittings from the GS of 9.8 and 27.3~meV, which are incompatible with our INS measurements. We were unable to find a CEF scheme with this GS configuration that fitted both the INS and magnetic susceptibility data.
\begin{table}[ht]
\caption{The parameters obtained from fitting $S_{\rm{mag}}(Q, \omega)$ from INS and magnetic susceptibility data.  $B_m^n$ were obtained from fitting the INS data. At 4 K, the value of $B_2^0$ was fixed whilst the other two CEF parameters were allowed to vary. The Lorentzian linewidths of the quasielastic scattering ($\Gamma_{\rm{QES}}$) and the first and second CEF excitations ($\Gamma_{\psi_2}$ and $\Gamma_{\psi_3}$) are also displayed.  The remaining parameters are obtained from fitting the magnetic susceptibility with anisotropic molecular field parameters ($\lambda_{ab}$ and $\lambda_0^c$) as well as temperature independent susceptibilities ($\chi_0^{ab}$ and  $\chi_0^c$).}
\label{CEFTab} 
\begin{ruledtabular}
 \begin{tabular}{c c c }

 & 4 K & 25 K \\
\hline
 $B_2^0$~meV& $-$0.61 & $-$0.61(4) \\
 $B_4^0$~meV & $-$0.013(3) & $-$0.007(2) \\
 $B_4^4$~meV & 0.412(8) & 0.463(8) \\
 $\Gamma_{\rm{QES}}$~(meV) & -- & 1.9(3) \\
$\Gamma_{\psi_2}$~(meV)& 2.5(2) & 1.6(3) \\
$\Gamma_{\psi_3}$~(meV) & 2.3(2) & 2.9(3) \\
 $\lambda_{ab}$ (mole/emu) & -- & $-$40.9 \\
 $\lambda_{c} $(mole/emu) & -- & $-$52.0 \\
 $\chi_0^{ab}$ ($\times 10^{-3}$ emu/mol) & -- & $-$0.404 \\
 $\chi_0^c$ ($\times 10^{-3}$ emu/mol) & -- & $-$1.936 \\

\end{tabular}
\end{ruledtabular}
\end{table}

We may now compare our results with those obtained from isostructural Ce$TX_3$ compounds. Like CeCoGe$_3$, the CEF model for CeRhGe$_3$ predicts a GS which is an admixture of  $\left|\pm \frac{5}{2} \right \rangle$ and $\left|\mp \frac{3}{2} \right \rangle$. \cite{CeRhGe32012} Both compounds have a significant  $B_4^4$, 0.463~meV for CeCoGe$_3$ and 0.294~meV for CeRhGe$_3$ which leads to this mixing. In CeRhGe$_3$, the  $\left|\pm \frac{3}{2} \right \rangle$ states are the largest components in the GS whilst for CeCoGe$_3$ it is $\left|\pm \frac{5}{2} \right \rangle$. In both compounds, the moments in the magnetically ordered state align along the $c$~axis. However $B_2^0$ is positive for CeRhGe$_3$ and a consideration of E$_a$ predicts a moment lying in the $ab$~plane. The alignment of the moment along $c$ is ascribed to two-ion anisotropic exchange interactions. Unlike CeCoGe$_3$, the easy axis of the magnetic susceptibility is in the $ab$~plane despite the moment alignment along $c$ below T$_{\rm{N}}$. The calculated value of $\langle \mu_z \rangle$ closely agrees to the result obtained from the magnetic neutron diffraction measurements and there is no evidence of a reduction of the cerium moment due to hybridization. In contrast to this, the CEF model for CeCoGe$_3$ correctly predicts the alignment of the ordered moment and the easy axis of the magnetic susceptibility.  However the observed moment is significantly reduced compared to the calculated value of $\langle \mu_z \rangle$. The reduction in moment is not as drastic as in the other pressure induced NCS CeRhSi$_3$ and CeIrSi$_3$. For example a CEF model of CeRhSi$_3$ \cite{CeRhSi3CEF} predicts a moment of 0.92~$\rm{\mu_B}$/Ce in the $ab$~plane whilst a moment of 0.12~$\rm{\mu_B}$/Ce in that direction is actually observed through neutron diffraction studies. \cite{CeRhSi3ND} This compound also has a very different magnetic structure, a spin-density wave with propagation vector (0.215,0,$\frac{1}{2}$). These results suggest that CeCoGe$_3$ has a degree of hybridization between that of CeRhGe$_3$ and CeRhSi$_3$.  This is consistent with the fact that CeRhSi$_3$ is closer to a QCP, having an onset of superconductivity at 1.2~GPa \cite{CeRhSi3SC} whilst CeCoGe$_3$ becomes superconducting at 5.5~GPa \cite{CeCoGe3SC} and CeRhGe$_3$ \cite{CeTX32008} does not become superconducting up to 8.0~GPa. The linewidths of the CEF excitations give an indication of the hybridization strength between the conduction electrons and the excited states. The linewidths obtained for CeCoGe$_3$ at 25~K were 1.6(3) and 2.9(3)~meV for transitions from the GS to $\psi_2$ and $\psi_3$ respectively. This is compared to values of 1.4(2) and 2.2(3)~meV obtained for CeRhGe$_3$. \cite{CeRhGe32012} The linewidth of the excitation to  $\psi_2$ was similar in both compounds whilst the excitation to  $\psi_3$ was broader in CeCoGe$_3$ than CeRhGe$_3$. However linewidths of 3.9(2) and 9.2(4)~meV were obtained for the CEF excitations of  CeRhSi$_3$ \cite{CeRhSi3ExpRep}, indicating stronger hybridization of all the states in the $J~=~\frac{5}{2}$ multiplet.

\section{Conclusions}
We have studied the magnetic ordering in CeCoGe$_3$ using single crystal neutron diffraction, inelastic neutron scattering, $\mu$SR and magnetic susceptibility. The transition to magnetic ordering below $T_{\rm{N1}}$ is observed with the emergence of oscillations in zero-field $\mu$SR spectra. We fitted the temperature dependence of the internal magnetic fields to a model of mean field magnet. Single crystal neutron diffraction measurements reveal magnetic ordering with a propagation vector  of \textbf{k} = (0,0,$\frac{1}{2}$) below $T_{\rm{N3}}$, \textbf{k} = (0,0,$\frac{5}{8}$) for $T_{\rm{N3}}~<~T~<~T_{\rm{N2}}$, and \textbf{k}~=~(0,0,$\frac{2}{3}$) for $T_{\rm{N2}}~<~T~<~T_{\rm{N1}}$. From a refinement of the integrated intensities we suggest a two-up, two-down magnetic structure below $T_{\rm{N3}}$ with moments of 0.405(5)~$\rm{\mu_B}$/Ce along the $c$~axis. Measurements of the (1~1~0) reflection indicate a ferromagnetic component between $T_{\rm{N3}}$ and $T_{\rm{N1}}$.  From this we suggest a two-up, one-down structure for the phase between  $T_{\rm{N2}}$ and $T_{\rm{N1}}$. INS measurements of polycrystalline CeCoGe$_3$ at low temperatures indicate two CEF excitations at 19 and 27~meV. At 4~K, we observe an additional peak at 4.5~meV due to spin wave excitations. Above $T_{\rm{N1}}$, this peak is not present but quasielastic scattering is observed. A linear fit to the temperature dependence of the quasielastic linewidth gives an estimate of $T_K~=~11(3)$~K. From an analysis of INS and magnetic susceptibility data with a CEF model, we propose a CEF scheme for CeCoGe$_3$. We are also able to account for the spin wave peak at 4.5~meV by the addition of an internal field along the $c$~axis. The CEF scheme correctly predicts the direction of the ordered moment but the observed magnetic moment at 2~K of 0.405(5)~$\rm{\mu_B}$/Ce is reduced compared to the predicted moment of 1.01~$\rm{\mu_B}$/Ce. We believe that the reduced moment is due to hybridization between the localized Ce$^{3+}$ f-electrons and the conduction band.  From considering the moment reduction, we deduce that CeCoGe$_3$ has a hybridization strength between that of the localized antiferromagnet CeRhGe$_3$ and the NCS CeRhSi$_3$. CeRhSi$_3$ exhibits SC at lower applied pressure than CeCoGe$_3$ whilst CeRhGe$_3$ does not exhibit SC up to at least 8.0~GPa. This is evidence for the important role of hybridization in the unconventional superconductivity of the Ce$TX_3$ series.

\begin{acknowledgments}
We acknowledge the EPSRC, UK for providing funding (grant number EP/I007210/1). DTA/ADH thank CMPC-STFC (grant number CMPC-09108) for financial support We thank T.E. Orton for technical support, S. York for compositional analysis and P. Manuel, B.D. Rainford and K.A. McEwen for interesting discussions. Some of the equipment used in this research at the University of Warwick was obtained through the Science City Advanced Materials: Creating and Characterising Next Generation Advanced Materials Project, with support from Advantage West Midlands (AWM) and part funded by the European Regional Development Fund (ERDF). 
\end{acknowledgments}

\bibliography{CeCoGe3_final}

\end{document}